\documentclass[reprint,prl,aps,superscriptaddress,showpacs,amsmath,amssymb]{revtex4-2}
\usepackage{graphicx}
\usepackage{bm}
\usepackage{hyperref}
\usepackage{cases}
\hypersetup{
	colorlinks=true,
	linkcolor=blue,
	citecolor=blue,
	urlcolor=blue,
}

\begin{document}

	\title{
	{Filtering one-way Einstein-Podolsky-Rosen steering}
}

\author{Ze-Yan~Hao}
\affiliation{CAS Key Laboratory of Quantum Information, University of Science and Technology of China, Hefei 230026, China}
\affiliation{CAS Center for Excellence in Quantum Information and Quantum Physics, University of Science and Technology of China, Hefei 230026, China}

\author{Yan~Wang}
\affiliation{CAS Key Laboratory of Quantum Information, University of Science and Technology of China, Hefei 230026, China}
\affiliation{CAS Center for Excellence in Quantum Information and Quantum Physics, University of Science and Technology of China, Hefei 230026, China}

\author{Jia-Kun~Li}
\affiliation{CAS Key Laboratory of Quantum Information, University of Science and Technology of China, Hefei 230026, China}
\affiliation{CAS Center for Excellence in Quantum Information and Quantum Physics, University of Science and Technology of China, Hefei 230026, China}

\author{Yu~Xiang}
\affiliation{State Key Laboratory for Mesoscopic Physics, School of Physics, Frontiers Science Center for Nano-optoelectronics, and Collaborative Innovation Center of Quantum Matter, Peking University, Beijing
	100871, China}
\affiliation{Collaborative Innovation Center of Extreme Optics, Shanxi University, Taiyuan, Shanxi 030006, China}

\author{Qiong-Yi~He}
\affiliation{State Key Laboratory for Mesoscopic Physics, School of Physics, Frontiers Science Center for Nano-optoelectronics, and Collaborative Innovation Center of Quantum Matter, Peking University, Beijing
	100871, China}
\affiliation{Collaborative Innovation Center of Extreme Optics, Shanxi University, Taiyuan, Shanxi 030006, China}	
\affiliation{Peking University Yangtze Delta Institute of Optoelectronics, Nantong, Jiangsu 226010, China}	

\author{Zheng-Hao~Liu}
\affiliation{CAS Key Laboratory of Quantum Information, University of Science and Technology of China, Hefei 230026, China}
\affiliation{CAS Center for Excellence in Quantum Information and Quantum Physics, University of Science and Technology of China, Hefei 230026, China}

\author{Mu~Yang}
\affiliation{CAS Key Laboratory of Quantum Information, University of Science and Technology of China, Hefei 230026, China}
\affiliation{CAS Center for Excellence in Quantum Information and Quantum Physics, University of Science and Technology of China, Hefei 230026, China}

\author{Kai~Sun}
\email{ksun678@ustc.edu.cn}
\affiliation{CAS Key Laboratory of Quantum Information, University of Science and Technology of China, Hefei 230026, China}
\affiliation{CAS Center for Excellence in Quantum Information and Quantum Physics, University of Science and Technology of China, Hefei 230026, China}

\author{Jin-Shi~Xu}
\email{jsxu@ustc.edu.cn}
\affiliation{CAS Key Laboratory of Quantum Information, University of Science and Technology of China, Hefei 230026, China}
\affiliation{CAS Center for Excellence in Quantum Information and Quantum Physics, University of Science and Technology of China, Hefei 230026, China}
\affiliation{Hefei National Laboratory, University of Science and Technology of China, Hefei 230088, China}

\author{Chuan-Feng~Li}
\email{cfli@ustc.edu.cn}
\affiliation{CAS Key Laboratory of Quantum Information, University of Science and Technology of China, Hefei 230026, China}
\affiliation{CAS Center for Excellence in Quantum Information and Quantum Physics, University of Science and Technology of China, Hefei 230026, China}
\affiliation{Hefei National Laboratory, University of Science and Technology of China, Hefei 230088, China}

\author{Guang-Can~Guo}
\affiliation{CAS Key Laboratory of Quantum Information, University of Science and Technology of China, Hefei 230026, China}
\affiliation{CAS Center for Excellence in Quantum Information and Quantum Physics, University of Science and Technology of China, Hefei 230026, China}
\affiliation{Hefei National Laboratory, University of Science and Technology of China, Hefei 230088, China}

\begin{abstract}
	
	\noindent{

		Einstein-Podolsky-Rosen (EPR) steering, a fundamental concept of quantum nonlocality, describes one observer's capability to remotely affect another distant observer's state by local measurements. Unlike quantum entanglement and Bell nonlocality, both associated with the symmetric quantum correlation, EPR steering depicts the unique asymmetric property of quantum nonlocality. With the local filter operation in which some system components are discarded, quantum nonlocality can be distilled to enhance the nonlocal correlation, and even the hidden nonlocality can be activated. However, asymmetric quantum nonlocality in the filter operation still lacks a well-rounded investigation, especially considering the discarded parts where quantum nonlocal correlations may still exist with probabilities. Here, in both theory and experiment, we investigate the effect of reusing the discarded particles from local filter. We observe all configurations of EPR steering simultaneously and other intriguing evolution of asymmetric quantum nonlocality, such as reversing the direction of one-way EPR steering. This work provides a perspective to answer ``{What is the essential role of utilizing quantum steering as a resource?}", and demonstrates a practical toolbox for manipulating asymmetric quantum systems with significant potential applications in quantum information tasks.		
		
	}
	
\end{abstract}	
\maketitle

\begin{figure*}[t]
	\centering
	\includegraphics[width=0.92\textwidth]{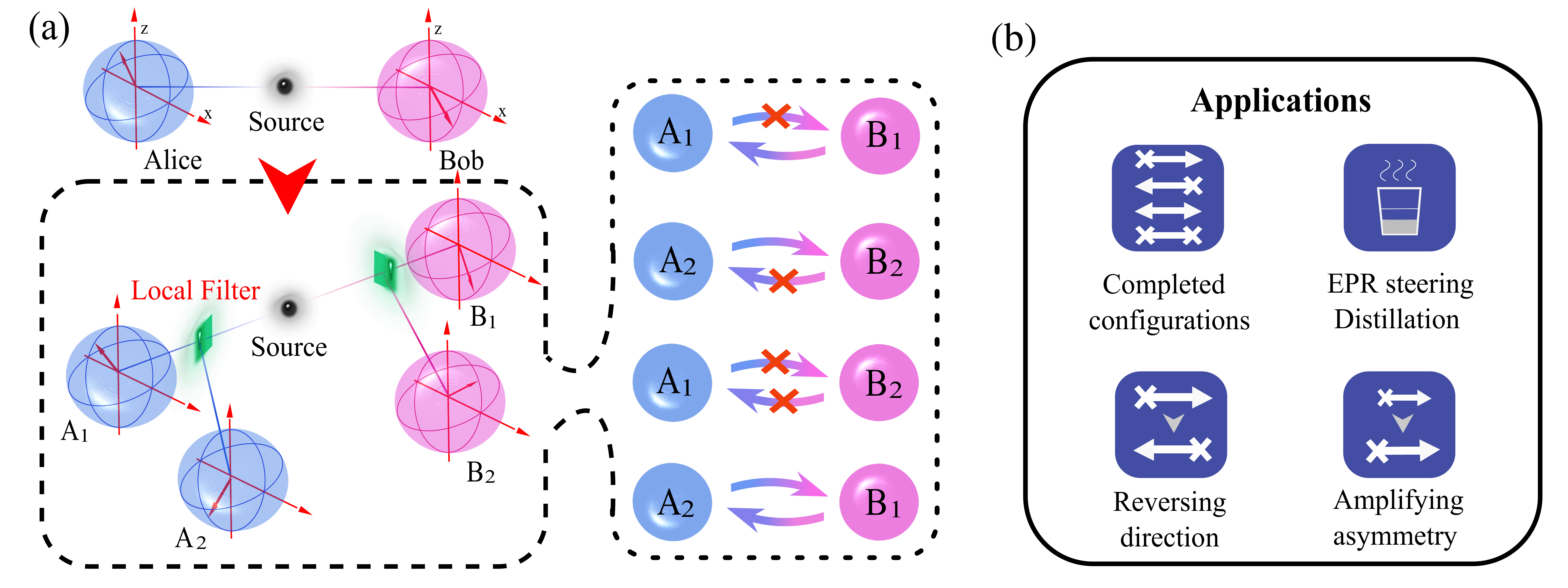}
	\caption { Illustration of EPR steering in the local filter operation. 
		(a).The local filter is performed with two output ports on each side of Alice and Bob, i.e., $A_1$ and $A_2$, $B_1$ and $B_2$. By using the well-chosen local filter, we can demonstrate different configurations of EPR steering with the states shared between $A_i$ and $B_j$, including  $A \rightarrow B$ one-way steering, $B \rightarrow A$ one-way steering, two-way steering and two-way unsteerable relations. 
		(b). With the effect of local filter operation, several scenarios of EPR steering could be observed, such as the exhibition of complete configurations of EPR steering, EPR steering distillation, reversal of the one-way EPR steering direction, and amplification of the capacity for one-way EPR steering.
	}\label{p0}
\end{figure*}

Quantum nonlocality, a cornerstone of quantum information studies, describes the phenomenon where measurement statistics among observers cannot be determined by local causality \cite{PhysicsPhysiqueFizika.1.195}. In the early studies of quantum nonlocality, symmetric correlations, such as entanglement \cite{horodecki2009quantum} and Bell nonlocality \cite{PhysRevLett.23.880,PhysRevLett.47.460,brunner2014bell}, were primarily investigated, in which the observers in the system have symmetric status. Einstein-Podolsky-Rosen (EPR) steering, a type of asymmetric quantum nonlocality, was originally put forward by Schrodinger \cite{schrodinger1935discussion,S1936} to argue the ``action at a distance" paradox \cite{einstein1935can}, and was reformulated by Wiseman $et\ al.$ in 2007 \cite{wiseman2007steering}. EPR steering characterizes the process that one observer can affect another observer's state by performing local measurements, and has drawn significant attention due to its extraordinary feature of asymmetric quantum nonlocality \cite{uola2020quantum,RevModPhys.95.011003}. Taking the bipartite system as an illustration, say Alice and Bob, it certifies that Alice can steer Bob's state when the assemblage of Bob's conditional states $\{\rho_{a|\vec{x}}\}$ after Alice's local measurements cannot be explained by a local hidden state (LHS) model \cite{wiseman2007steering,uola2020quantum}, i.e., the following equation,
\begin{equation}
\rho_{a|\vec{x}}=\int P(a | \vec{x}, \lambda) p_{\lambda} \rho_{\lambda}\mathrm{d}\lambda,
\end{equation}
where $\vec{x}$ is Alice's measurement setting with the output $a\in \{0,1\}$, $p_\lambda$ is the distribution of LHS $\rho_\lambda$  parameterized by variable $\lambda$, and $P(a|\vec{x}, \lambda)$  corresponds to the output probability distribution of Alice measurements. According to asymmetric status of Alice and Bob, EPR steering reveals the unique directional property that could lead to one-way nonlocality \cite{handchen2012observation,PhysRevLett.112.200402,PhysRevLett.114.060403,PhysRevLett.114.060404,sun2016experimental,PhysRevLett.116.160403,xiao2017demonstration,tischler2018conclusive}; that is, Alice could steer Bob's state, but Bob cannot steer Alice's state. Based on the directional characteristic, all asymmetric configurations in the bipartite system include two-way steerable, one-way steerable ($A \rightarrow B$ and $B \rightarrow A$), and two-way unsteerable scenarios \cite{quintino2015inequivalence}. Due to the unique directional property, EPR steering has been widely discussed for use in many quantum information protocols, such as one-side device-independent quantum-key distribution (QKD) \cite{branciard2012one,RN39,RN38,gallego2015resource}, subchannel discrimination \cite{PhysRevLett.114.060404,RN18},  multipartite EPR steering constrained by monogamy relations \cite{PhysRevLett.122.070402,armstrong2015multipartite,doi:10.1126/science.aao2254,hao2022demonstrating}, and quantum network based on directional quantum correlations \cite{wang2020deterministic,PhysRevLett.127.170405}.

In the practical protocol of the quantum nonlocal scenario, the local filter, which discards some components of the system in local environments \cite{gisin1996hidden} and could be represented in the form of the Kraus operator \cite{hirsch2016entanglement}, is exploited as a toolbox to revive the nonlocality in a system originally no-violating Bell inequality. Since proposed by Gisin \cite{gisin1996hidden}, the local filter operation has attracted continuous attention due to its wide applications in the purification of entanglement \cite{PhysRevLett.78.574,kwiat2001experimental,wang2006experimental} and the activation of hidden nonlocality \cite{masanes2008all,jones2020exploring}, and has been exploited in many quantum tasks, such as quantum repeater \cite{zhao2003experimental} and QKD \cite{mishra2020increasing}. Recently, local filter operation has also been applied in EPR steering scenarios to distillate EPR steering \cite{nery2020distillation,ku2022complete,RN16} or reveal hidden steerability \cite{pramanik2019revealing}. However, a comprehensive investigation of performing filters on EPR steering, which may provide novel views to understand the quantum asymmetric property, still lacks a thorough discussion. Especially in a conventional process including the local filter operation, some discarded parts may still exist quantum nonlocal correlations that fail to pass the filter \cite{zhou2020purification}, which can also be an important resource for building the quantum network \cite{wang2020deterministic,PhysRevLett.127.170405}.

Here, we theoretically and experimentally study the detailed effect of performing local filter operation on the manipulation of EPR steering by using all output components of filter operations on both sides of Alice and Bob, as shown in Fig. \ref{p0}(a). With different combinations of two sides' outputs, the essential configurations of EPR steering can be verified.
For example, for an original one-way EPR steering from Alice to Bob, i.e., $A \rightarrow B$ one-way steering, with the help of local filter operations, we can reverse the direction of one-way steering to implement the $B \rightarrow A$ one-way steering, In addition to the observation of the above-newfangled scenario, in the presence of local filter operations on both sides, we realize EPR steering distillation which enhances the steerability, and amplifies the asymmetry of one-way EPR steering in which the steerability of $A \rightarrow B$ one-way steering becomes larger while $B \rightarrow A$ unsteerability becomes more unsteerable. Such an amplification could lead to a simpler one-way EPR steering confirmation.
Our work contributes to a practical toolbox to flexibly manipulate asymmetric quantum nonlocality, which helps understand quantum nonlocality profoundly and has potential in secure QKD \cite{RN17} and construction of quantum internet \cite{kimble2008quantum}.\\

In this work, we consider the local filter operation acting on a two-qubit system, which can be written in the following form: \cite{gisin1996hidden,kwiat2001experimental,pramanik2019revealing},
\begin{equation}
F_{A_1}=\left(\begin{array}{cc}
a_1 & 0 \\
0 & a_2
\end{array}\right),\ F_{B_1}=\left(\begin{array}{ll}
b_1 & 0 \\
0 & b_2
\end{array}\right). \label{E1}
\end{equation}
For the initial state $\rho_{A B}$, the local filter operation contributes to the evolution of $(F_{A_1} \otimes F_{B_1}) \rho_{A B} (F_{A_1} \otimes F_{B_1})^{\dagger} / \mathrm{Tr}((F_{A_1} \otimes F_{B_1})\rho_{A B} (F_{A_1} \otimes F_{B_1})^{\dagger}) $. After transmission, many components of the system are discarded which may still possess quantum nonlocal correlations \cite{zhou2020purification}. Here, we utilize all the discarded components as shown in Fig \ref{p0}(a) (reflection part) to simultaneously reveal additional configurations of EPR steering and expand the applications of the local filter operation. By exploiting the filter described in Eq. \eqref{E1} and the corresponded reflection part depicted in Fig \ref{p0}(a), we obtain an ensemble of local filters ${F_{A_1},\ F_{B_1},\ F_{A_2},\ F_{B_2}}$, which satisfy the condition $F_{A_1}^\dagger F_{A_1}+F_{A_2}^\dagger F_{A_2}=I$ and $F_{B_1}^\dagger F_{B_1}+F_{B_2}^\dagger F_{B_2}=I$, where the usage of the ensemble brings no particle loss.

We apply the asymmetric two-qubit states as the initial input, which  can demonstrate $A \rightarrow B$ one-way steering \cite{xiao2017demonstration},
\begin{equation}
\rho_{A B}=\eta |\Phi(\theta)\rangle\langle\Phi(\theta)|+(1-\eta)I_{A} / 2 \otimes \rho_{B}^{\theta}, \label{rho}
\end{equation}
where
$
|\Phi(\theta)\rangle=\cos \theta|00\rangle+\sin \theta|11\rangle
$, $I_{A}$ is identity matrix and $\rho_{B}^{\theta}=\mathrm{T r_{A}}(|\Phi(\theta)\rangle\langle\Phi(\theta)|)$. {Steering from $A$ to $B$ can be observed within the range of $1/\sqrt{3}<\eta \leq 1/{\sqrt{1+2 \sin ^2(2 \theta)}}$ in a scenario involving three measurement settings. However, as the number of settings increases to infinity, Bob is unable to steer Alice when $\cos ^2(2 \theta)\geq(2 \eta-1)/{(2-\eta) \eta^3}$. } After the evolution of local filters $F_{ij}=F_{A_i}\otimes F_{B_j}$, the EPR steering source transforms to
\begin{equation}
\rho_{A_i B_j}=\frac{F_{ij} \rho_{A B} F_{ij}^{\dagger} }
{\mathrm{Tr}(F_{ij} \rho_{A B} F_{ij}^{\dagger}) } \label{pro}
\end{equation}
with the probability $\mathrm{Tr}(F_{ij} \rho_{A B} F_{ij}^{\dagger})$. As shown in Fig. \ref{p0}(a),
with the well-chosen local filters, we can demonstrate complete configurations of EPR steering. Furthermore, as depicted in Eq. \eqref{pro}, the process exhibits intriguing possibilities for exploring various applications within local filter tasks, as illustrated in Fig. \ref{p0}(b).		

\begin{figure}[h]
	\centering
	\includegraphics[width=0.48\textwidth]{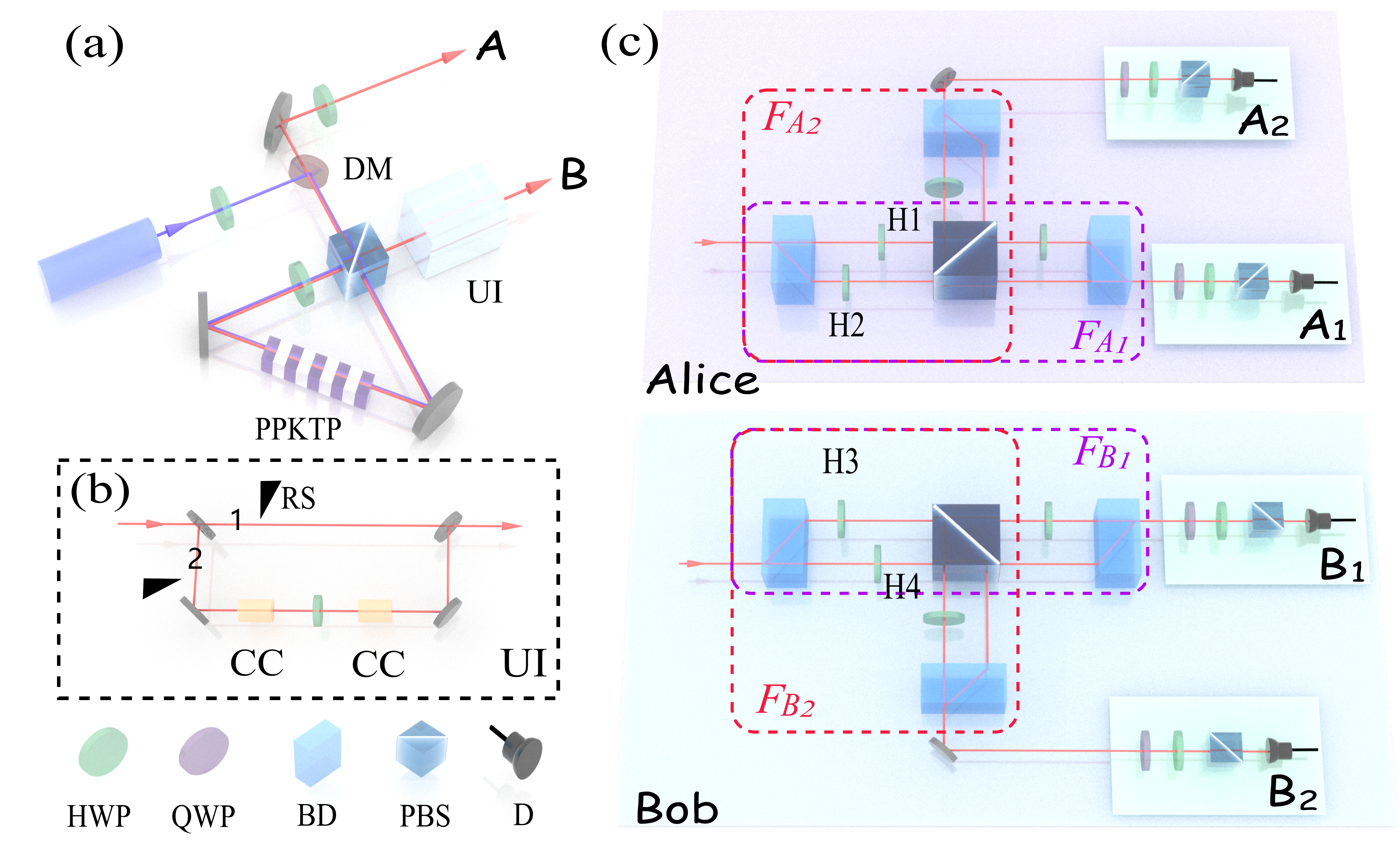}
	\caption{  Experimental preparation of states and local filter operation in the optical setup. (a). The source of polarization-entangled states, which is generated by the spontaneous parametric down-conversion process. (b). The unbalanced interfermeter (UI), which is exploited to mix different components to generate mixed states in Eq. \eqref{rho}. (c). The optical process of local filter operations. Taking Alice's side as an example, the initial photons that injured the filters are divided into two optical paths with different polarizations through the beam displacer (BD). We adjusted the transitivity and reflectivity of polarization states in two paths by half wave plates (HWP) and a polarization beam splitters (PBS) to construct the local filter operations in Eq. \eqref{E1}. Another BD is used to combine two paths and erase the path information. The measurement apparatus is composed of polarization analyzers and detectors.
	}\label{p2}
\end{figure}

{\it Experimental setup---} We generate the state in Eq. \eqref{rho} in an optical setup with encoding polarizations  $|H\rangle $ as $|0\rangle$, and $|V\rangle$ as $|1\rangle$ ($|H\rangle$ and $|V\rangle$ correspond to horizontal and vertical polarizations, respectively). The details of source are shown in the Appendix (A.3). Since the source is polarization entangled, the polarization dependent loss is employed to construct local filters as shown in Fig. \ref{p2}(c). Taking the local filter of $F_{A_1}$ as an example, by using a beam displacer (BD), which splits the input beam into two orthogonally polarized beams, the polarization components $\left| H\right\rangle$ and $\left| V\right\rangle$ are separated. Half-wave plates (HWP) H1 and H2 are used to adjust the polarization information, which decide the weight of transmission and reflection after polarization beam splitters (PBS). The parameters of the local filter in Eq. \eqref{E1} are determined by the degree of H1 and H2. Then, another BD is employed to combine two polarization components into one path. Similarly,  $F_{B_1}$ is decided by the degrees of H3 and H4 (more details are in Appendix), where the two-qubit operation $F_{11}$ is expressed as $F_{A_1} \otimes F_{B_1}$. Compared to the previous work discarding the part filtered out by $F_{11}$ \cite{pramanik2019revealing,ku2022complete}, we have no photon loss in our experimental process since the local filters $F_{A_2}$, $F_{B_2}$ are constructed to manufacture the complete ensemble of local operations.

The measurement devices of every observer include a quarter-wave plate (QWP), an HWP, and a PBS to analyze the polarization of photons. Before sending signals to the coincidence counting, we use two 3 nm bandwidth interference filters to remove the stray light and gather the photons with single-photon detectors.

\begin{figure*}[t]
	\centering
	\includegraphics[width=0.98\textwidth]{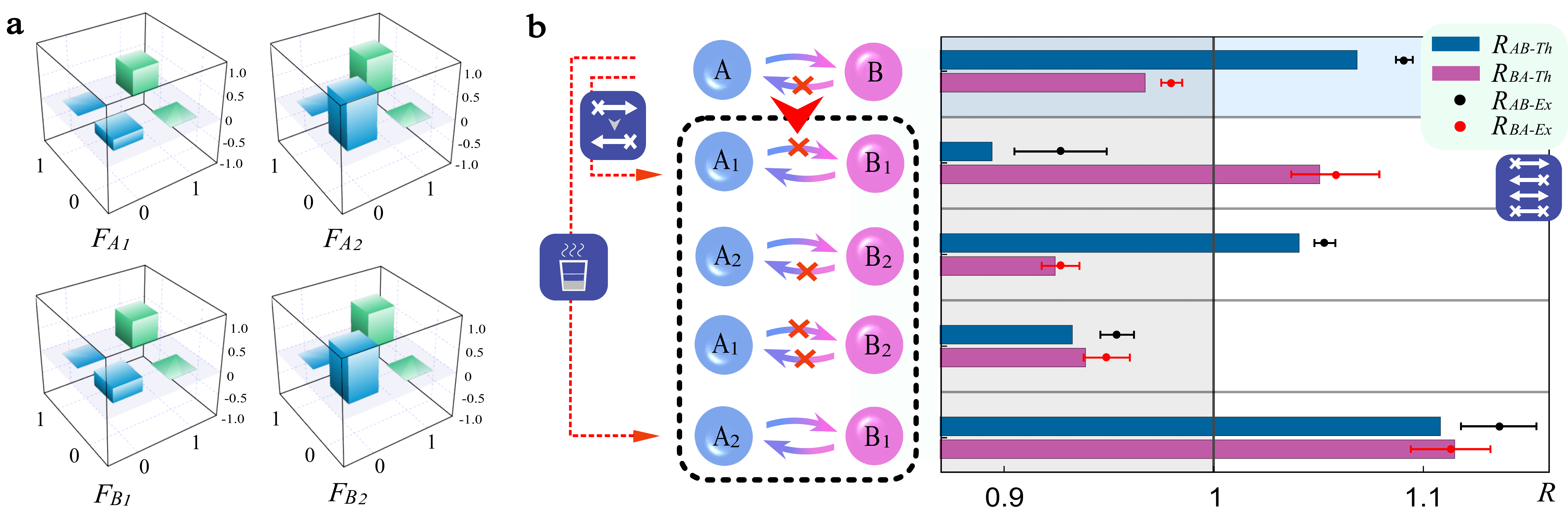}
	\caption{ Process of filtering one-way EPR steering. (a). The process tomography of the ensemble of local filter operations, including $F_{A_1}$, $F_{A_2}$ of Alice, and $F_{B_1}$, $F_{B_2}$ of Bob. (b). The results demonstrate complete configurations of EPR steering with an initially asymmetric state $\rho_{AB}$. We employ steering radius to quantify the steerability, the white region ($R>1$) indicates steerability and the gray region ($R<1$) corresponds to unsteerability. The blue and purple columns represent the theoretical results of $R_{A \rightarrow B}$ and $R_{B \rightarrow A}$, and the black and red dots represent the experimental results. It is noteworthy that the reversal of the direction of one-way steering can be achieved when filtering  $\rho_{AB}$ to $\rho_{A_1B_1}$. In the evolution of $\rho_{A_2B_1}$, the steerability is distilled in both directions.}\label{p3}
\end{figure*}

{\it Experimental Results---} We exploit the steering radius $R_{A \rightarrow B}$ and $R_{B \rightarrow A}$ to quantify EPR steering in this work (see in Appendix). {$R_{A \rightarrow B}>1$ indicates that Alice can steer Bob's state, while $R_{A \rightarrow B}<1$ indicates that Alice cannot steer Bob's state under three projective measurement settings. The exploiting of steering radius can guarantee the sufficiency and necessity of steering verification in this scenario.} We prepare the initial state $\rho_{AB}$ with parameters $\theta=0.452(3)$, $\eta=0.647(3)$ in Eq. \eqref{rho}, achieving a fidelity of $0.9959(6)$. Through criterion with three measurements settings, we obtain $R_{A \rightarrow B}=1.091(4)$ and $R_{B \rightarrow A}=0.980(5)$, revealing that Alice can steer Bob's state, while Bob cannot steer Alice's state. To fully explore the potential of the discarded photons, we perform process tomography on the ensemble of local filter operations ($F_{A_1}$, $F_{A_2}$, $F_{B_1}$, and $F_{B_2}$), achieving an average fidelity of 0.962(1) (see Fig. \ref{p3}(a) and more details are in Appendix). The states resulting from the application of local filters exhibit complete configurations of EPR steering, including A $\rightarrow$ B one-way steering, B $\rightarrow$ A one-way steering, unsteerability for both directions and two-way steering as shown in Fig. \ref{p3}(b).

\begin{figure*}[t]
	\centering
	\includegraphics[width=0.95\textwidth]{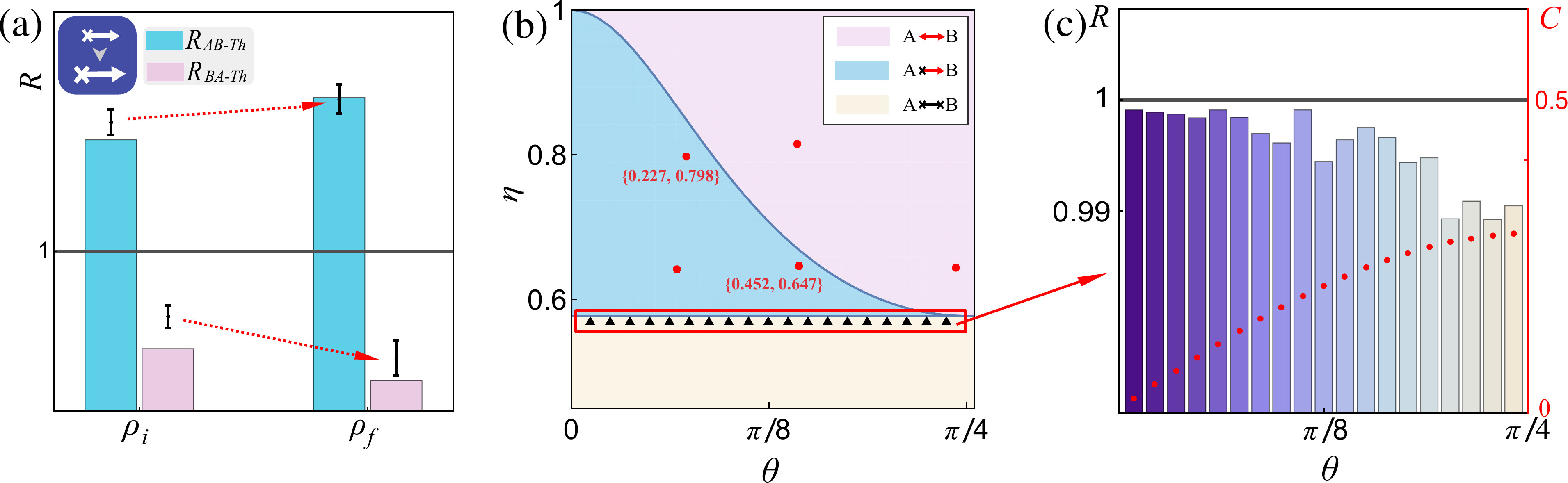}
	\caption{ Amplification of the asymmetry for one-way EPR steering and demonstration of the entanglement states without hidden EPR steering. (a). By exploiting the local filter operation, the one-way steering can be amplified in the evolution of the initial state $\rho_i$ transforming to the final state $\rho_f$, where the $A \rightarrow B$ steerability is amplified, and the  $B \rightarrow A$ steerability is attenuated. (b). The distribution of states with different parameters. The pink, blue and yellow regions represent two-way EPR steering,  A $\rightarrow$ B one-way steering and two-way unsteering of initial states, respectively. {The red dots represent experimental initial states (details are shown in Appendix), where states labeled by parameters correspond to state in Fig. 3 and Fig. 4(a)}, and the black triangle represents the states exploited for numerical analysis. (c). The numerical simulation results of hidden EPR steering in unsteerable systems. The column and red dots indicate the steering radius and the concurrence of initial states. The results show that the initial states are entangled, where concurrence larger than 0 ($C>0$), but cannot be distilled to be steerable by local filter operations.
	}\label{p4}
\end{figure*}

With the above results, we can notice several interesting applications except demonstrating complete configurations of EPR steering. As shown in Fig. \ref{p3} (b), when filtering $\rho_{A B}$ to $\rho_{A_1 B_1}$, the direction of one-way steering is reversed, where one-way steering of A $\rightarrow$ B turns to that of  B $\rightarrow$ A. It is noteworthy that B $\rightarrow$ A one-way steering is originally impossible to be observed in the type of states in Eq. \eqref{rho} {(as shown in Fig. \ref{p4}(b)}. When filtering $\rho_{A B}$ to $\rho_{A_2 B_1}$, local operation leads to the steerability of both directions being purified. This demonstrates the presence of hidden  B $\rightarrow$ A EPR steering within the initial state, where Bob can not steer Alice and can be recovered via EPR steering distillation. The error bars in the figures are obtained by Poissonian counting statistics. The differences among the error bars of the output ports rely on the probabilities in Eq. \eqref{pro} which reflects the detection of coincidence counting between different outputs $A_i$ and $B_j$.

Moreover, as shown in Fig. \ref{p4}(a), we investigate another initial state parameterized by $\theta=0.227(2)$, $\eta=0.798(1)$ passing through an ensemble of local filters owning $\{a_1,a_2\}=\{0.70,0.20\}$, $\{b_1,b_2\}=\{0.12,0.16\}$, our results indicate that the asymmetry of one-way steering can be amplified by only local operation. The $A \rightarrow B$ steerability is enhanced while $B \rightarrow A$ steerability is attenuated. This amplification results in a reinforced gap between the steerability from A to B and the unsteerability from B to A, simplifying the task of detecting one-way steering.

Furthermore, we theoretically and numerically analyze the existence of hidden EPR steering in different initial states (more theoretical details are in Appendix). We discover that the hidden EPR steering disappears when the initial state in Eq. \eqref{rho} is unsteerable in both directions. For nineteen initial states with $\eta$ smaller than the bound $1/\sqrt{3}$ of two-way unsteerability as shown in Fig. \ref{p4}(b), we randomly choose 40000 local filter operations and calculate the maximal steering radius of the corresponding final states. The numerical results are shown in Fig. \ref{p4}(c), in which none of the states after local filter operations can demonstrate EPR steering in either direction. The results indicate that the type of states in Eq. \eqref{rho} are entangled but without hidden EPR steering when they are two-way unsteerable, which demonstrates that the steering-unsteering boundary elegantly defining the purified-unpurified boundary.\\

{\it Conclusions---}In summary, in this work, all configurations of EPR steering, including two-way steering, $A \rightarrow B$ one-way EPR steering, $B\rightarrow A$ one-way EPR steering,  and two-way unsteerable correlations are observed simultaneously with different outputs of local filter operations on both Alice's and Bob's sides. 
In the view of quantum resources, configurations of quantum resources after filtering one-way EPR steering can provide abundant practical sources to construct a steering network \cite{PhysRevLett.127.170405}, and multiuser communication network \cite{wang2020deterministic,fan2022robust}. {Remarkably, by combining different outputs to efficiently reuse discarded resources, we have observed that almost all filtering processes are asymmetric in nature, except the balanced situation $F_{A_1}=F_{A_2}=F_{B_1}=F_{B_2}$. Within this plethora of asymmetric processes, the majority of the reused resources can only be characterized by asymmetric correlations, such as EPR steering. As a result, we can address a crucial question:{\it ``What is the essential role of utilizing quantum steering as a resource?''} The answer is that: to maximize the utility of the discarded components from a distillation process, the asymmetric evolution of process is inevitable, and it is imperative design tasks built upon the foundation of asymmetric quantum correlations.}

We also develop the practical operations employed in EPR steering systems. The direction of one-way steering is reversed via a local operation, exhibiting that $A \rightarrow B$ one-way EPR steering can transform to $B \rightarrow A$ one-way EPR steering, {where $B \rightarrow A$ one-way EPR steering is an initially impossible correlation for state in Eq. \eqref{rho} as shown in Fig. \ref{p4}(b). This result is also a direct demonstration of that local filters are treated to be a quantum operations not just a process in state preparation.} The reversal of the EPR steering direction can be applied to a network containing a quantum server \cite{wang2020deterministic}, where the quantum server can redispose the network structure via only filtering operations. We also show that the local filters can be used to distillate one-way EPR steering to two-way, i.e. the no-cloning condition to guarantee the secure teleportation \cite{PhysRevLett.115.180502}, and the distillation of EPR steering has already  been used to increase the secret-key rate of QKD \cite{RN21,RN16}. We also demonstrate that employing local filters can amplify the asymmetry of one-way steering and significantly facilitate detecting one-way EPR steering. The result reveals the asymmetric influence of a local operation acting on an EPR steering system, which can amplify one side's steerability while attenuating the other side.

Moreover, by theoretically and numerically calculating the options of local filters to optimize output states having the most steerable ability, we show that there is no hidden EPR steering in our two-way unsteerable initial states. This conclusion may provide a fundamental view of research on hidden nonlocality since the result helps to search the bound of whether hidden nonlocality can be activated. Answering this question has significance in the research of bound entanglement \cite{horodecki1998mixed,tendick2020activation}, which is a long-term open question of quantum theory \cite{horodecki2022five}. It would also be interesting to extend this framework to multipartite and high dimension seniors to underpin the understanding of hidden nonlocality \cite{tendick2020activation}.

The aforementioned applications provide flexible and efficient operations in EPR steering to underpin the capability of manipulating quantum nonlocality in practical applications. The filtering explores the capability of employing full properties of EPR steering in quantum information tasks, which provides a novel viewpoint to understand the distinctive property of EPR steering and can be a useful and powerful instrument in future quantum theory development to be the critical operation solving the open problem of EPR steering studies \cite{uola2020quantum,PRXQuantum.3.030102}. \\

\begin{acknowledgments}
This work was supported by the Innovation Program for Quantum Science and Technology (Nos. 
2021ZD0301200,
2021ZD0301400), National Natural Science Foundation of China (Nos.
11821404,
61975195,
61725504, U19A2075,
12125402, 12004011), Anhui Initiative in Quantum Information Technologies (No. 
AHY060300), the Fundamental Research Funds for the Central Universities (No.
WK2030380017),
USTC Research Funds of the Double First-Class Initiative (No. YD2030002024). 
\end{acknowledgments}

\bibliographystyle{plain}

\newpage

\appendix
\section{Appendix}
\subsection {A.1 Steering radius}
 To quantify EPR steering, we exploit the steering radius $R_{A \rightarrow B}$ and $R_{B \rightarrow A}$ \cite{sun2016experimental,xiao2017demonstration}. Taking Alice to Bob as an example, after Alice makes the local measurement on directions $\{\vec{n}_1,\ \vec{n}_2,\ \vec{n}_3\}$, Bob can obtain several condition states explained by an assembly of LHS $\{ p_i\rho_i \}$. The maximum value of the radius of the Bloch vector in ${ \rho_i}$ exceeding $1$ indicates the unphysicality of corresponding LHS. Thus, we define the minimum radius based on the measurement directions as $ r_{\left\{\vec{n}_{1}, \vec{n}_{2}, \vec{n}_{3}\right\}}=\min _{\left\{p_{i} \rho_{i}\right\}}\left\{\max \left\{\left|\vec{R}_{i}\right|\right\}\right\}$ where $>1$ indicates the nonexistent of LHS. Considering the optimal measurement directions, the steering radius $R_{A \rightarrow B}=\max _{\left\{\vec{n}_{1}, \vec{n}_{2}, \vec{n}_{3}\right\}}\left\{r_{\left\{\vec{n}_{1}, \vec{n}_{2}, \vec{n}_{3}\right\}}\right\}$
possesses the ability to sufficiently and necessarily quantify the steerability under three projective measurement settings. 

In an experiment, to obtain the steering radius of $A \rightarrow B$,  bob's assemblage of conditional states $\{p_{a|\vec{n}}\rho_{a|\vec{n}}\}$ is necessary. In the situation of three measurement settings, 6 components of $\{p_{a|\vec{n}}\}$ and $\rho_{a|\vec{n}}$ are included in the assemblage which can be obtained in experimental measurement, and a LHS containing 8 comments $\{ p_i\rho_i \}$ is sufficient to construct Bob's experimental conditional assemblage \cite{xiao2017demonstration}. The unphysicality of LHS, also can be determined as Bloch vector $\max \{|\vec{R_i}|\}$ of $\rho_i$ $>1$, is then reveal the steerability. {With the construction method of assemblage of conditional state, the theoretical witness can be performed to analysis experimental data, where a cost function can be determined. 
\begin{equation}
\begin{aligned}
F_{p_i,\rho_i}=\\ \sum_{a|\vec{n}} ((\sum_i p_i & \rho_i p(i,a|\vec{n}) -p_{a|\vec{n}}\rho_{a|\vec{n}})^2+ (\sum_i p_i p(i,a|\vec{n}) -p_{a|\vec{n}})^2)
\end{aligned}
\end{equation}
Therefore, a semi-definite program (SDP) can help us to calculate the error between LHS model and experiment data.}

\begin{equation}
\begin{aligned}
E_{|\vec{R_\lambda}|}=\min _{p_i,\rho_i} F_{p_i,\rho_i}\\
\text { s.t. } & Vec(\rho_i)< |\vec{R_\lambda}|\quad  \sum_i p_i=1 \quad  \forall i,
\end{aligned}
\end{equation}
And, another SDP can help us to calculate the steering radius.

\begin{equation}
\begin{aligned}
R_{A \rightarrow B}=\max _\lambda |\vec{R_\lambda}|
\\
\text { s.t. } & E_{|\vec{R_\lambda}|}<err \quad \forall \lambda,
\end{aligned}
\end{equation}
where, $err$ is the limitation of the error of steering radius. We set $err$ as $0.000012$ to guarantee the average error of constructed radius less than 0.001. $R_{A \rightarrow B}>1$ indicates that Alice can steer Bob's state, while $R_{A \rightarrow B}<1$ indicates that Alice cannot steer Bob's state in the situation of three measurement settings.

\begin{figure*}[t]
	\centering
	\includegraphics[width=0.93\textwidth]{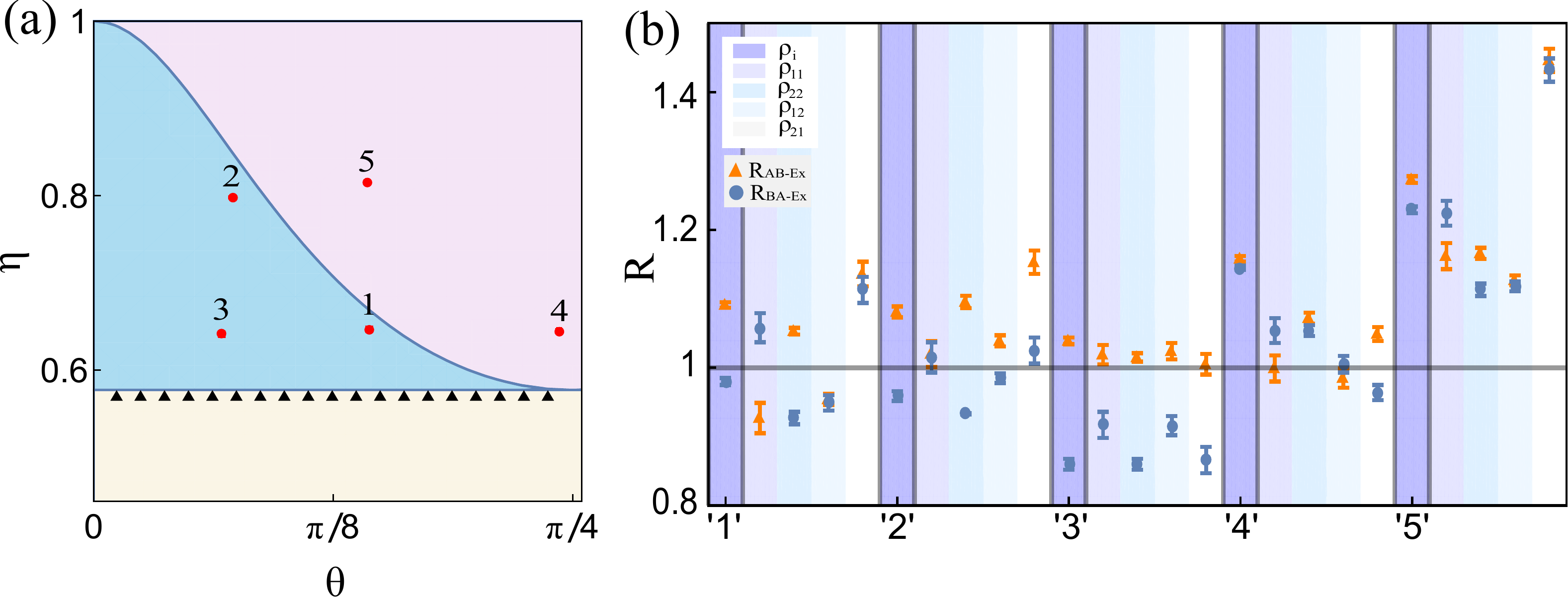}
	\caption{ The results of five initial states acting by local filter operations. (a) The location of states with different parameters. (b) The detection of EPR steering after local filter operations, where the orange triangles indicate the steering radius of $A \rightarrow B$ ($R>1$ means EPR steering), and the blue dots indicate the steering radius of $B \rightarrow A$. The initial states are shown in blue columns.}\label{all}
\end{figure*}

\subsection{A.2 The states without hidden EPR steering}
In this section, we discuss the question about hidden EPR steering, which describes the concept of detecting the steerabillity after evolution. In the main text, we have demonstrated the numerical results that indicate the disappearance of hidden EPR steering when the initial state is unsteerable in both directions (with the condition $\eta<\sqrt{3}/3$ \cite{xiao2017demonstration}). Here, we exhibit the theoretical derivation of the aforementioned conclusion. According to the proof in \cite{hirsch2016entanglement}, for the initial state:
\begin{equation}
\rho_{A B}=\eta |\Phi(\theta)\rangle\langle\Phi(\theta)|+(1-\eta)I_{A} / 2 \otimes \rho_{B}^{\theta}, \label{rho2}
\end{equation}
 $\rho_{AB}$ is unsteerable from Bob to Alice if and only if the state:
\begin{equation}
\rho_{F_B}=\frac{I \otimes F_{B} \rho_{W}(\eta) I \otimes F_B^\dagger}{\operatorname{Tr}\left(I \otimes F_{B} \rho_{W}(\eta) I \otimes F_B^\dagger\right)} \label{rhoF}
\end{equation}
is unsteerable from Bob to Alice for arbitrary local filters $F_B$, where $\rho_{W}(\eta)$ is the Werner state:
\begin{equation}
\rho_{W}(\eta)=\eta\left|\phi^{+}\right\rangle\left\langle\phi^{+}\right|+(1-\eta) I / 4,
\end{equation}
and $\left|\phi^{+}\right\rangle=(|00\rangle+|11\rangle) / \sqrt{2}$. In our scenario of two-way unsteerability for $\rho_{AB}$, $\rho_{F_B}$ is apparently $B \rightarrow A$ unsteerable (Werner states are unsteerable when $\eta<\sqrt{3}/3$ in the three setting measurements \cite{zhen2016certifying}). 

Therefore, after Bob applying his local operation $F_B'$, the state
\begin{equation}
\rho_{AB}'=\frac{I \otimes F_{B}' \rho_{AB} I \otimes F_B'^\dagger}{\operatorname{Tr}\left(I \otimes F_{B}' \rho_{AB} I \otimes F_B'^\dagger\right)}
\end{equation}
remains unsteerable from Bob to Alice, since the state  
\begin{equation}
\rho_{F_B}'=\frac{I \otimes (F_{B}F_{B}') \rho_{W}(\eta) I \otimes (F_B'F_B)^\dagger}{\operatorname{Tr}\left(I \otimes (F_{B}F_{B}') \rho_{W}(\eta) I \otimes (F_B'F_B)^\dagger\right)} \label{E7}
\end{equation}
also takes the form of Eq. \eqref{rhoF}. Next, Alice applies her local operation $F_A'$, resulting in the state transforming to its final form:
\begin{equation}
\rho_{AB_f}=\frac{F_A' \otimes F_{B}' \rho_{AB} F_A' \otimes F_B'^\dagger}{\operatorname{Tr}\left(F_A' \otimes F_{B}' \rho_{AB} F_A' \otimes F_B'^\dagger\right)}.
\end{equation}
According to Eq. \eqref{E7}, $\rho_{AB_f}$ is unsteerable from Alice to Bob when the state:
\begin{equation}
\rho_{F_B}'=\frac{F_A' \otimes (F_{B}F_{B}') \rho_{W}(\eta) F_A' \otimes (F_B'F_B)^\dagger}{\operatorname{Tr}\left(F_A' \otimes (F_{B}F_{B}') \rho_{W}(\eta) F_A' \otimes (F_B'F_B)^\dagger\right)} \label{E9}
\end{equation} is unsteerable from Bob to Alice \cite{hirsch2016entanglement}. We define the operation $F_B''=F_B F_B'$, and Eq. \eqref{E9} takes the form:
\begin{equation}
\rho_{F_B}'=\frac{F_A' \otimes (F_{B}'') \rho_{W}(\eta) F_A' \otimes (F_B'')^\dagger}{\operatorname{Tr}\left(F_A' \otimes (F_{B}'') \rho_{W}(\eta) F_A' \otimes (F_B'')^\dagger\right)},
\end{equation}
which has been proven in \cite{hirsch2016entanglement} to be unsteerable form  $A$ to $B$ for arbitrary local filter operations when the state $\rho_{A B}=\eta |\Phi(\theta)\rangle\langle\Phi(\theta)|+(1-\eta)\rho_{A}^{\theta} \otimes I_B/2$ is unsteerable with the parameter $\eta<\sqrt{3}/3$. Based on the proof in \cite{quintino2015inequivalence}, it is established that any local filters applied by Alice on Bob's state cannot convert the state from unsteerable in the $B \rightarrow A$ direction to steerable. Consequently, we derive the conclusion that the state $\rho_{AB_f}$ remains two-way unsteerable when the initial state $\rho_{AB}$ is two-way unsteerable. Thus, we have successfully demonstrated that the unsteerable states described in Eq. \eqref{rho} do not possess hidden EPR steering.

\subsection{A.3 Preparation of states} As shown in Fig. \ref{p2}(a), a 404 nm continuous-wave diode laser with the full width at half maximum (FWHM) of the spectra 0.05 nm is applied to pump the entanglement source. A 20 mm type-II cut nonlinear crystal of periodically poled $\mathrm{KTiOPO_4}$(PPKTP) is placed in a polarization Sagnac interferometer \cite{fedrizzi2007wavelength}, where the spontaneous parametric down-conversion process leads the input photon $|H\rangle$ transforming to output two photons $|HV\rangle$. The initial state $\cos \theta|H\rangle+\sin \theta|V\rangle$ prior to injection into the Sagnac interferometer then transforms to a polarization-entangled state $\cos \theta|HV\rangle+\sin \theta|VH\rangle$, where $\theta$ is decided by a half wave plate (HWP). A $45^{\circ}$ HWP is placed to flip the polarization, leading to the generation of the entangled state $|\Psi(\theta)\rangle = \cos \theta|HH\rangle + \sin\theta|VV\rangle$. When $\theta=45^{\circ}$, we obtain the maximally entangled state, and the brightness is approximately $4000~\mathrm{pairs/(s)} $ with 0.9mW power of the pump laser.

\begin{figure*}[t]
	\centering
	\includegraphics[width=0.93\textwidth]{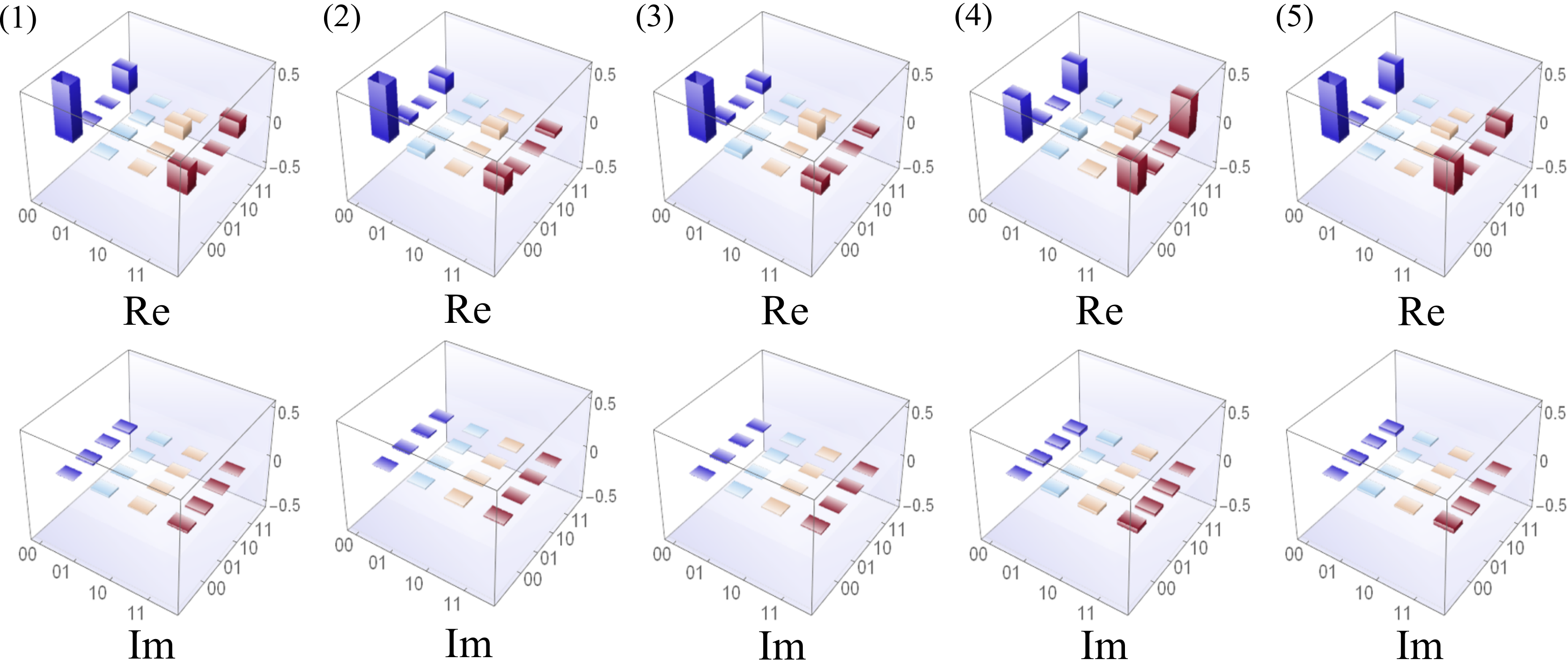}
	\caption{ Real and imaginary parts of the states reconstructed via tomography. The labels (1)-(5) correspond to the label numbers of states in Fig. \ref{all}. }\label{tomo}
\end{figure*}

To engineer the target mixed state in Eq. (3) in the main text, an unbalanced interferometer composed of two paths in Fig. \ref{p2}(b) is applied. In path-1, no operation interacts to the photon, the outside state is still $|\Psi(\theta)\rangle=\cos \theta|HH\rangle+\sin \theta|VV\rangle$. In path-2, the first birefringent crystal (BC) provides decoherence between $H$ and $V$ polarization to make the state become $\cos^2 \theta|HV\rangle \langle HV| +\sin^2 \theta|VH\rangle \langle VH|$. A $22.5^{\circ}$ HWP provides evolution $|H\rangle \rightarrow 1/\sqrt{2} (|H\rangle+|V\rangle) $, $|V\rangle \rightarrow 1/\sqrt{2} (|H\rangle-|V\rangle) $, and the photon after second BC finally becomes $I_{A} / 2 \otimes \rho_{B}^{\theta}$. In addition, removable shutters (RS) are employed to control the brightness in each path accurately, allowing for precise preparation of the desired states described in Eq. (3) of the main text.

We prepared five experimental states to investigate the effects of local filter operations on them, and the results are shown in Fig. \ref{all}. {Deciding which parameters of initial states and local filter operations need to be achieved in the experiment is determined by prior numerical calculations.} In Fig. \ref{all}(a), the pink, blue and yellow regions correspond to two-way steering, $A \rightarrow B$ one-way steering, and two-way unsteering, respectively. The red dots represent the five experimental states, labeled with numbers beside them. The results in Fig. \ref{all}(b) demonstrate the effect of local filters acting on EPR steering, which concludes by demonstrating complete configurations, reversing the direction of one-way steering, EPR steering distillation, and amplifying asymmetry of one-way steering. The tomography of the initial states is shown in Fig. \ref{tomo}, where the labels of the states correspond to the numbers assigned to the states in Fig. \ref{all}(a).

\subsection{A.4 Controlling of local filter parameters} In our optical system, the parameters of the local filter are determined by H1, H2, H3, and H4 in Fig. \ref{p2}(c). Taking $F_{A_1}$ as an example: initially, the first BD is applied to split the polarization components $|H\rangle$ and $|V\rangle$ into different paths. By Rotating H1 to $\alpha_1$, the state $|V\rangle$ can be adjusted to $\sin{2\alpha_1}|H\rangle-\cos{2\alpha_1}|V\rangle$, where the components $\sin{2\alpha_1}|H\rangle$ can pass the PBS and turn to $\sin{2\alpha_1}|V\rangle$ via an HWP set at $45^{\circ}$. Similarly, Rotating H2 to $\alpha_2$ leads $|H\rangle$ transforming to $\cos{2\alpha_2}|H\rangle$. After recombining the two paths with another BD, the process of the local filter $F_{A_1}$ is completed, and it can be expressed as:
\begin{equation}
F_{A_1}=\left(\begin{array}{cc}
\cos{2\alpha_2} & 0 \\
0 & \sin{2\alpha_1}
\end{array}\right).
\end{equation}
Moreover, $F_{B_1}$, $F_{A_2}$, and $F_{B_2}$ are in the form of:
\begin{equation}
\begin{aligned}
F_{B_1}=\left(\begin{array}{cc}
\cos{2\alpha_3} & 0 \\
0 & \sin{2\alpha_4}
\end{array}\right), \\
F_{A_2}=\left(\begin{array}{cc}
\sin{2\alpha_2} & 0 \\
0 & \cos{2\alpha_1}
\end{array}\right), \\
F_{B_2}=\left(\begin{array}{cc}
\sin{2\alpha_3} & 0 \\
0 & \cos{2\alpha_4}
\end{array}\right),
\end{aligned}
\end{equation}	
where $\alpha_1$,  $\alpha_2$, $\alpha_3$, and $\alpha_4$  represent the rotation angles corresponding to H1, H2, H3, and H4, respectively.

\subsection{A.5 Process tomography of local filter operation}  In this work, we experimentally implement several local filter operations using an optical setup. These filters are designed using polarization-dependent loss to control the components of $H$ and $V$ polarization. To better manipulate and observe the process of local filter operations, we employ process tomography.

Taking local filters of Alice ($F_{A_1}$ and $F_{A_2}$) as an example, $F_{A_1}$ and $F_{A_2}$ satisfy $F_{A_1}^{\dagger}F_{A_1} + F_{A_2}^{\dagger}F_{A_2} = 1$. The combination of $F_{A_1}$ and $F_{A_2}$ determines a completely positive map $\mathcal{E}$, with the input state $\rho$ in the following output form:
\begin{equation}
F_{A_1}\rho F_{A_1}^{\dagger} + F_{A_2}\rho F_{A_2}^{\dagger} = \mathcal{E}(\rho) = \sum_{m, n=0}^{3} \chi_{m n} \hat{\sigma}_{m} \rho \hat{\sigma}_{n}^{\dagger}, \label{EE6}
\end{equation}
where $\hat{\sigma}_{m}$ is the Pauli basis in our two-qubit situation, and $\chi_{m n}$ completely describes the process. The positive map in Eq. \eqref{EE6} can be easily achieved by employing the method described in \cite{PhysRevLett.93.080502}.

In a specific measurement with projector $|x\rangle$ and the injected state $\rho$, the counting numbers of measurements on outputs $A_1$ and $A_2$ determine the normalization parameters in this process:
\begin{equation}
\left\{
\begin{aligned}
N_x(A_1) = N\langle x | F_{A_1}\rho F_{A_1}^{\dagger} | x\rangle \\
N_x(A_2) = N\langle x | F_{A_2}\rho F_{A_2}^{\dagger} | x\rangle \\
N_x(A_1) + N_x(A_2) = N\langle x| \mathcal{E}(\rho) & |x\rangle, \label{E6}
\end{aligned}
\right.
\end{equation}
where $N$ is the total number of photons, and $N_x(A_1)$ and $N_x(A_2)$ are the counting numbers of output ports $A_1$ and $A_2$.

We express the local filter operations $F_{A_1}$ and $F_{A_2}$ in the Pauli basis as follows:
\begin{equation}
F_{A_1} = \sum_{m=0}^{3} a_{1m} \hat{\sigma}_{m}, \ F_{A_2} = \sum_{m=0}^{3} a_{2m} \hat{\sigma}_{m}.
\end{equation}

Ultimately, we can derive the concrete form of the local filter operations as shown in Fig. 3(a) in the main text according to the following equations:
\begin{equation}
\left\{
\begin{aligned}
\sum_{m, n=0}^{3} (a_{1m}a_{1n}+a_{2m}a_{2n}) \hat{\sigma}_{m} \rho \hat{\sigma}_{n}^{\dagger} = \sum_{m, n=0}^{3} \chi_{m n} \hat{\sigma}_{m} \rho \hat{\sigma}_{n}^{\dagger}\\
\mathrm{Tr}(F_{1}^{\dagger}F_{1}) = N_t(A_1)/N, \ \ \ \mathrm{Tr}(F_{2}^{\dagger}F_{2}) = N_t(A_2)/N,
\end{aligned}
\right.
\end{equation}
where $N_t(A_1)$ and $N_t(A_2)$ are the total numbers of photons in ports $A_1$ and $A_2$. The process to solve Eq. (21) employs a semi-definite program (SDP).

\begin{equation}
\begin{aligned}
\min_{a_{1},a_{2}} \sum_{m, n=0}^{3} (a_{1m}a_{1n}+a_{2m}a_{2n}-\chi_{m n})^2 \\
\text { s.t. } \mathrm{Tr}(F_{1}^{\dagger}F_{1}) = N_t(A_1)/N  \mathrm{Tr}(F_{2}^{\dagger}F_{2}) & = N_t(A_2)/N  \\  \forall a_1, a_2. \qquad \qquad \qquad \qquad \qquad \qquad \ \ &
\end{aligned}
\end{equation}

The errors in the work are obtained using Poissonian counting statistics.

\end{document}